\documentclass[traditabstract]{aa} 
\usepackage{graphicx}
\usepackage{natbib}
\usepackage{color}

\bibpunct{(}{)}{;}{a}{}{,} 

\usepackage{txfonts}

\begin{document}
\title{{The blue-edge problem of the V1093 Her instability strip revisited using evolutionary models with atomic diffusion}}
\titlerunning{The blue-edge problem of the V1093 instability strip revisited}

\author{S.~Bloemen\inst{\ref{KUL},\ref{RUN},\ref{KITP}} \and Haili~Hu\inst{\ref{KITP},\ref{SRON}} \and C.~Aerts\inst{\ref{KUL},\ref{RUN},\ref{KITP}}  \and M.~A.~Dupret\inst{\ref{ULG}}  \and R.~H.~\O stensen\inst{\ref{KUL}} \and P.~Degroote\inst{\ref{KUL}} \and E.~M\"uller-Ringat\inst{\ref{TUB}} \and T.~Rauch\inst{\ref{TUB}}}
\institute{{Instituut voor Sterrenkunde, KU Leuven, Celestijnenlaan 200D, B-3001 Heverlee, Belgium \label{KUL}} \and
{Department of Astrophysics, IMAPP, Radboud University Nijmegen, PO Box 9010, NL-6500 GL Nijmegen, The Netherlands\\
\email{s.bloemen@astro.ru.nl}\label{RUN}} \and
{Kavli Institute for Theoretical Physics, Kohn Hall, University of California, Santa Barbara, CA 93106, USA\label{KITP}}\and
{SRON, Netherlands Institute for Space Research, Sorbonnelaan 2, NL-3584 CA Utrecht, The Netherlands\label{SRON}} \and
{Institut d'Astrophysique et de G\'eophysique de l'Universit\'e de Li\`ege, All\'ee du 6 Ao\^ut 17, 4000 Li\`ege, Belgium \label{ULG}} \and
{Institute for Astronomy and Astrophysics, Kepler Center for Astro and Particle Physics, Eberhard Karls University, Sand 1, D-72076 T\"ubingen, Germany\label{TUB}}}

\date{Received 20 December 2013 / Accepted 2 September 2014}
\abstract{We have computed a new grid of evolutionary subdwarf B star (sdB) models from the start of central He burning, taking into account atomic diffusion due to radiative levitation, gravitational settling, concentration diffusion, and thermal diffusion. We have computed the non-adiabatic pulsation properties of the models and present the predicted $p$-mode and $g$-mode instability strips.  In previous studies of the sdB instability strips, artificial abundance enhancements of Fe and Ni were introduced in the pulsation driving layers. In our models, the abundance enhancements of Fe and Ni occur naturally, eradicating the need to use artificial enhancements. We find that the abundance increases of Fe and Ni were previously underestimated and show that the instability strip predicted by our simulations solves the so-called blue edge problem of the subdwarf B star g-mode instability strip. The hottest known $g$-mode pulsator, KIC\,10139564, now resides well within the instability strip {even when only modes with low spherical degrees ($l \leq 2$) are considered.}
}

\maketitle

\section{Introduction and state of the art}

Subdwarf B stars (sdBs) are extreme horizontal branch stars burning helium in their cores \citep[for a review on sdBs, see][]{Heber2009}. They typically have masses around 0.5$\,M_\odot$ and have only a thin hydrogen layer left at their surfaces, which is too thin to sustain hydrogen shell burning. Because of their thin envelopes, sdBs are unusually bright in the near-UV. They are identified as the source of the unexpected UV-upturn observed in elliptical galaxies \citep{PodsiadlowskiHan2008}. The formation of sdBs is not well understood. It is unclear how the sdB progenitors lost most of their hydrogen envelopes. 

More than 50 per cent of the sdBs have been found to be part of short period binaries \citep{MaxtedHeber2001}. This suggests that binary evolution may play an important role in their formation. Several evolutionary channels have been proposed: (1) common-envelope ejection channels, leading to sdBs in binaries with orbital periods below $\sim 10\,$d and a white dwarf or low-mass main sequence companion; (2) a stable Roche lobe overflow channel resulting in binary systems with orbital periods between $\sim  10\,$d and $\sim 100\,$d and much thicker hydrogen envelopes; and (3) a double helium white dwarf merger channel that gives rise to single sdB stars with very thin hydrogen envelopes and a wider mass range \citep{HanPodsiadlowski2002,HanPodsiadlowski2003}. Recently, sdBs with planetary companions were found \citep{SilvottiSchuh2007,CharpinetFontaine2011, BeuermannDreizler2012}, rejuvenating the idea that interactions of a star with a planetary companion might also lead to the formation of sdBs \citep{SokerHarpaz2000}. The evolutionary scenarios remain largely untested. Since they result in different predicted populations of sdBs, accurate determinations of their masses and envelopes, as well as periods of sdB binaries, can be used to discriminate between the different channels. 

Many sdBs show stellar oscillations. Pulsation modes for which the pressure force is the dominant restoring force, $p$-modes, were predicted in sdB stars by \citet{CharpinetFontaine1996} and discovered around the same time by \citet{KilkennyKoen1997}. The first detection of gravity ($g$-) mode pulsations, for which buoyancy is the dominant restoring force, in sdBs was accomplished by \citet{GreenFontaine2003}. Asteroseismic techniques can therefore be used to probe the internal structure of sdBs and to measure the masses of their cores and envelopes {(see e.g. \citealt{Van-GrootelCharpinet2010}, \citealt{FontaineBrassard2012} and references therein, and \citealt{Van-GrootelCharpinet2013})}. Asteroseismology of sdBs is not only useful to constrain the evolutionary scenarios, but also allows us to test our knowledge of  physical processes such as atomic diffusion and stellar winds {\citep{CharpinetFontaine2009, HuTout2011}}, as well as to test tidal theory through spin-orbit synchronizations of sdBs in short period binaries ({\citealt{Van-GrootelCharpinet2008,CharpinetVan-Grootel2008}}; \citealt{PabloKawaler2011,PabloKawaler2012}).
Recent reviews of sdB asteroseismology can be found in \citet{FontaineBrassard2006}, \citet{Ostensen2009,Ostensen2010}, {\citet{CharpinetBrassard2009}}, \citet{Kawaler2010} and {\citet{CharpinetVan-Grootel2013}}.

Pulsations in sdBs are driven by the opacity ($\kappa$) mechanism, which operates in stellar layers where chemical elements are partially ionized {\citep{UnnoOsaki1989}. \citet{CharpinetFontaine1996} showed that $p$-modes could be driven if radiative levitation could bring enough iron in the driving region. \citet{CharpinetFontaine1997} confirmed this idea by implementing iron abundance profiles into static sdB models, which were based on radiative levitation calculations assuming diffusive equilibrium. Using time-dependent diffusion calculations, \citet{FontaineBrassard2006a} showed that the convergence towards diffusive equilibrium is fast compared to the evolutionary timescale, thereby validating this diffusive equilibrium approach.  \citet{FontaineBrassard2003} argued that the same opacity mechanism was responsible for the driving of $g$-modes and recognized that radiative levitation is a crucial factor for the mechanism to work in this case as well.   In this pioneering study, the stars that were theoretically predicted to show $g$-mode pulsations were several thousand Kelvin too cool compared to the observed pulsators.} Using OP \citep{BadnellBautista2005} instead of OPAL \citep{IglesiasRogers1996} opacity tables, and by enhancing not only the Fe, but also the Ni abundance in the envelope, \citet{JefferySaio2006} obtained a predicted instability strip that reduced the so-called blue-edge problem. 

Using artificial enhancements of Fe and Ni in the pulsation driving region, \citet{HuNelemans2009} showed that gravitational settling, thermal diffusion and concentration diffusion acting on H and He shifts the theoretical blue-edge of the instability strip farther to about 1000\,K from the observed value. Recently, \citet{HuTout2011} added diffusion due to radiative levitation to their evolution code, next to the already implemented gravitational settling, thermal diffusion and concentration diffusion, and solved the diffusion equations for H, He, C, N, O, Ne, Mg, Fe and Ni. Because of the inclusion of radiative levitation, Fe and Ni enhancements are built up in the pulsation driving region, therefore eliminating the need to include artificial abundance enhancements. The authors showed that the resulting evolutionary models can excite low-degree $g$-modes at relatively high effective temperatures, suggesting that the blue-edge problem could be resolved. In this work, we present a grid of models computed using the codes of \citet{HuTout2011} to readdress the instability strip issue.

\section{Model grid}
We have set up tools to compute grids of evolutionary tracks of sdB stars using a modified version of the stellar evolution code \texttt{STARS} \citep{Eggleton1971}. The most important changes are the implementation of gravitational settling, thermal diffusion and concentration diffusion by \citet{HuGlebbeek2010}, the implementation of radiative levitation by \citet{HuTout2011} and the coupling of the \texttt{STARS} code to the non-adiabatic pulsation code \texttt{MAD} \citep{Dupret2001} as described in \citet{HuDupret2008}.
The evolutionary tracks are computed starting from red giant branch models from which most of the envelope was stripped off. The free parameters of the starting models are the total mass of the star and the mass of the remaining hydrogen envelope. For the grid presented here, we assumed solar metallicity with the metal mixture of \citet{GrevesseNoels1993} and no mass loss due to winds during the sdB evolution. The total masses ($M_*$) range from 0.35\,$M_\odot$ to 0.55\,$M_\odot$ and the envelope masses from $10^{-5}\,M_*$ to $10^{-1.8}\,M_*$. The grid contains over nine thousand models on 77 evolutionary tracks. The computation of one evolutionary track takes about 6 days of CPU time on a single core of a 2.8GHz quad-core Intel Xeon X5560 Nehalem CPU.

\section{Instability strips}

\subsection{Iron and nickel abundance enhancements}

\begin{figure}
\includegraphics[width=88mm]{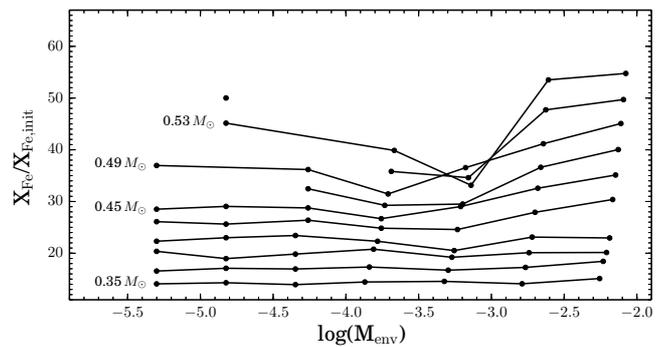}
 \caption{Abundance enhancement of Fe in the driving region at $T = 200\,000$\,K, $10^7$ years after the start of central He burning. The lines connect the enhancement factors of models with the same total mass but different envelope masses. }
  \label{sdbgrid_FIG_FeEnhance}
\end{figure}

\begin{figure}
\includegraphics[width=88mm]{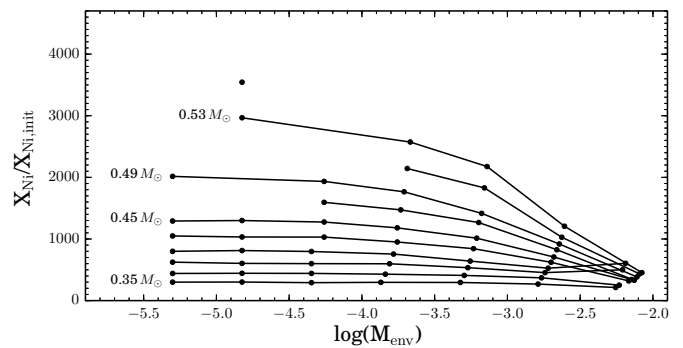}
 \caption{Same as Fig.\ \ref{sdbgrid_FIG_FeEnhance} but for Ni instead of Fe. In earlier work, enhancement factors of 10 or 20 were assumed, which is clearly an underestimation.}
  \label{sdbgrid_FIG_NiEnhance}
\end{figure}

\citet{HuTout2011} showed that the build up of the Fe abundance in the driving region proceeds relatively quickly and reaches an equilibrium situation in about $10^5$ years, as was also shown by \citet{FontaineBrassard2006a}. \citet{HuTout2011} found that it takes longer before the equilibrium situation is reached for Ni, but this also happens in less than $10^7$\,yr. For the test track presented in that paper, it was  clear that the Ni abundance increase compared to the initial abundance was much higher than was previously assumed. From our grid spanning a range of envelope masses and total masses, we have measured the height of the Fe and Ni abundance peaks in the driving region $10^7$ years after the start of the sdB evolution on the zero-age horizontal branch. The resulting abundance enhancement factors compared to the initial abundance are shown in Fig.\ \ref{sdbgrid_FIG_FeEnhance} for Fe and in Fig.\ \ref{sdbgrid_FIG_NiEnhance} for Ni.  {For the same core mass, a higher envelope mass results in a higher effective temperature and a higher surface gravity. The former increases the efficiency of the radiative levitation, but this is counteracted by the increased settling due to the higher surface gravity. The net result is that models with different envelope masses, but the same core mass, tend to have similar abundance enhancements. The figures only show a datapoint for the evolutionary tracks that reached an age of $10^7$ years, which is not the case for all tracks because of numerical convergence problems.}

\begin{figure}
\includegraphics[width=88mm]{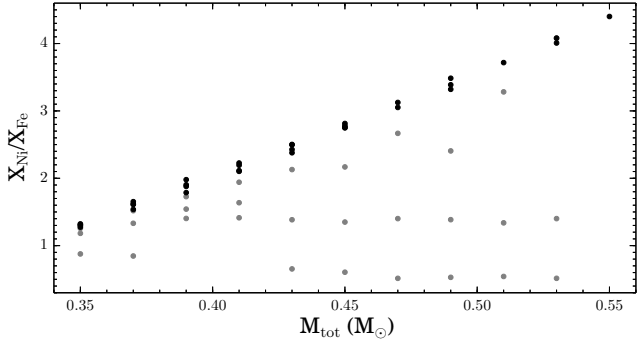}
 \caption{{ Ratio of the abundance of Ni and Fe in the driving region at $10^7$ years after the start of the central He burning as a function of the total mass of the star. Black dots show models with $\log (M_{\rm env}/M_{\odot})<3.5$ and grey dots models with  $\log (M_{\rm env}/M_{\odot})\geq3.5$. }}
  \label{sdbgrid_FIG_FevsNi}
\end{figure}

While previous studies of the instability strips of sdBs used parametrized enhancement factors of up to 20 for the abundances of Fe and Ni in the driving regions \citep{JefferySaio2006,HuDupret2008}, our simulations predict that especially the Ni abundance increase is orders of magnitude higher. For low mass models ($M_*=0.35\,M_{\odot}$), we find an increase by a factor of $\sim 300$ for Ni. For higher mass models, we find enhancement factors of up to 4000. {The large difference in enhancement factors between Fe and Ni is caused by the initial underabundance of Ni compared to Fe of a factor of 16.1 in the \citet{GrevesseNoels1993} mixture. The absolute Ni abundances in the driving region are typically between 1 and 4 times that of Fe (see Fig.\ \ref{sdbgrid_FIG_FevsNi}), which is compatible with the results presented by \citet[][see right panel of their Figure 5]{MichaudRicher2011}. For models with a low envelope mass ($\log (M_{\rm env}/M_{\odot})<3.5$ shown in black), there is a clear trend of an overabundance of Ni compared to Fe that increases with the total mass of the star.}
 
\subsection{Pressure and gravity mode instability strips}\label{SEC_instabstrips}

\begin{figure}
\includegraphics[width=88mm]{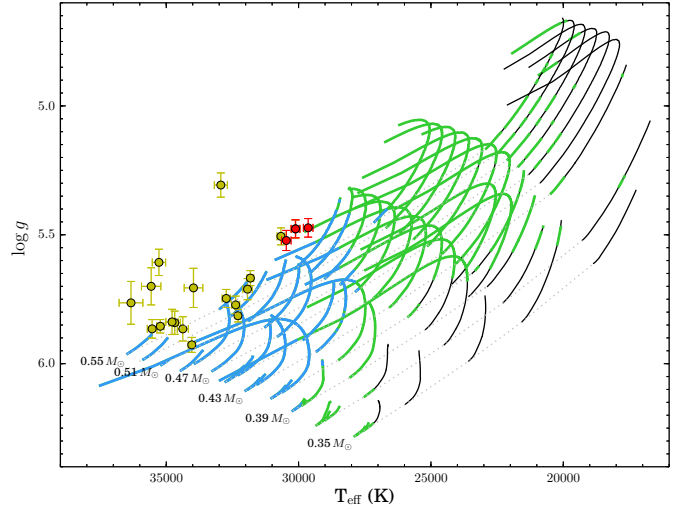}
 \caption{ Instability strip for $p$-mode pulsations in sdB stars. The initial points of evolutionary tracks of stars with identical core masses are connected with a grey dotted line and the masses are indicated on the plot. Envelope masses rise from left to right and range from $10^{-5}\,M_*$ to $10^{-1.8}\,M_*$. Sections of the tracks where more than 1 (10) $p$-modes with spherical degree $l\leq 3$ are predicted to be excited, are coloured in green (blue). The rest of the tracks is shown in black. A sample of known $p$-mode pulsators are indicated with yellow points and red points indicate known hybrid pulsators.}
  \label{sdbgrid_FIG_pmodeInstab}
\end{figure}

\begin{figure}
\sidecaption
 \includegraphics[width=88mm]{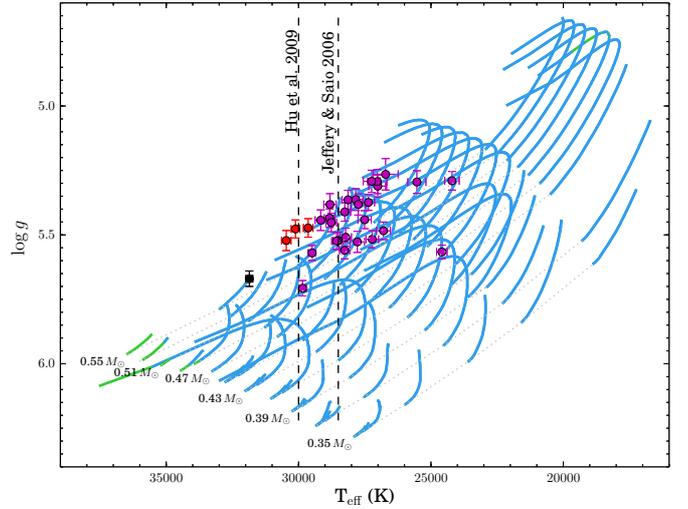}
 \caption{ Same as Fig.\ \ref{sdbgrid_FIG_pmodeInstab} but for $g$-mode pulsations in sdB stars. Red points still indicate known hybrid pulsators that show $p$- and $g$-modes. Magenta points indicate known $g$-mode pulsators. The black point indicates KIC\,10139564 \citep{BaranReed2012}, which is a hybrid pulsator that predominantly shows $p$-mode pulsations. The vertical dashed lines indicate the blue edges of the instability strips predicted by \citet{JefferySaio2006} and \citet{HuNelemans2009}.}
  \label{sdbgrid_FIG_gmodeInstab}
\end{figure}

The higher Fe and Ni  opacity bumps in the driving region influence the instability strips. In Fig.\ \ref{sdbgrid_FIG_pmodeInstab} we show the predicted instability strip for $p$-mode pulsations. The evolutionary tracks are coloured green (blue) when {at least 1 (10)} $p$-modes of spherical degree $l\leq 3$ have a positive work integral in our non-adiabatic pulsation computations. Figure \ref{sdbgrid_FIG_gmodeInstab} shows the same for $g$-modes. The dots indicate a sample of pulsators taken from \citet{GreenFontaine2008} that was already shown in \citet{Ostensen2010}: $p$-mode pulsators are shown in yellow, $g$-mode pulsators in magenta and hybrid pulsators that show both $p$- and $g$-mode pulsations in red. {The small apparent offset between the spectroscopically determined surface gravities and effective temperatures, and the evolutionary tracks for typical $\sim 0.47 M_\odot$ sdBs was seen in earlier studies (see e.g. Fig.~1 in \citealt{Ostensen2009} where sdB models from \citealt{KawalerHostler2005} are shown). As shown in \citet{DormanRood1993}, the exact position of the zero-age extreme horizontal branch depends not only on the core mass of the sdBs but also on the assumed metallicity, and a change from solar to subsolar metallicity shifts the zero-age extreme horizontal branch to lower $\log g$ and higher $T_{\rm eff}$.}

All pulsators fall within the instability strips, which in our simulations extend to slightly cooler temperatures than the known pulsators in the case of $p$-modes and to hotter temperatures in the case of $g$-modes. The predicted instability strips for $g$-mode pulsations, including $l=1,2$ and $3$ modes, showed a blue edge around 28\,500\,K in the study by \citet{JefferySaio2006}, and around 30\,000\,K in the study by \citet{HuNelemans2009}. { In our case, we find that $g$-modes can be excited at all temperatures covered by the grid, and the models with several ($\geq 10$) excited modes have temperatures above $\sim 33\,500\,$K.
}

\begin{figure*}
\sidecaption
  \includegraphics[width=170mm, trim=45 0 90 0, clip]{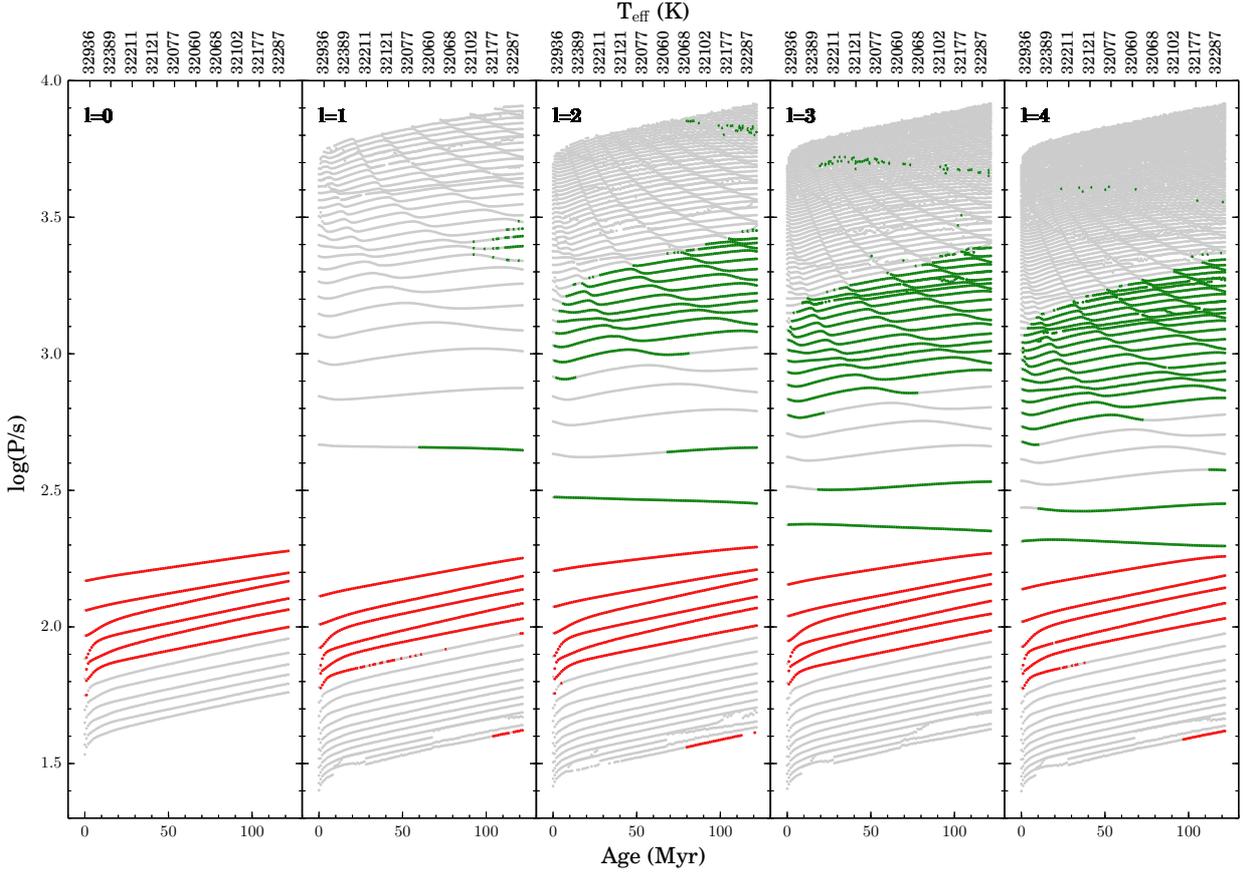}
 \caption{{Predicted pulsation spectra per spherical degree for the evolutionary track with a total mass of $0.47$M$_\odot$ and an envelope mass of $9.0\ 10^{-5}$M$_\odot$. Excited $p$-modes are shown in red, excited $g$-modes in green. The effective temperature of the models is shown on the top axis.}}
  \label{fig_0.47pgmodeVSl}
\end{figure*}

{As an illustration, Fig.\ \ref{fig_0.47pgmodeVSl} shows which modes are driven along the evolutionary track of an sdB with a  total mass of $0.47$M$_\odot$ and an envelope mass of $9.0\ 10^{-5}$M$_\odot$. \citet{FontaineBrassard2003} showed that the blue-edge of the instability strip is highly sensitive to the spherical degree. This can also be seen in this example, in which $g$-modes of degree $l=2,\,3$ and 4 are excited at all ages (except for the very first model), while the $l=1$ $g$-modes are not excited during the first $~50$ Myr, and then only 1 mode is excited until almost the end of the evolutionary track. The example also illustrates that modes of high spherical degrees are not needed to explain the observed pulsators: plenty of $l=2$ $g$-modes are found to be excited at $\sim 32\,000\,$K. Such low-degree modes are easier to observe in photometry since they suffer much less from partial geometric cancellation.
More details on the model, including the effective temperature, the surface gravity and the Ni and Fe abundance enhancements in the driving region as a function of time, are given in two tables in Appendix \ref{ch_appendix_047}. The pulsational characteristics of all models in the grid are available in electronic form at \texttt{https://fys.kuleuven.be/ster/Projects/sdbgrid/}.
}

\subsection{Envelope mixing and KIC\,10139564, the hottest hybrid sdB star pulsator}

\begin{figure}
\sidecaption
 \includegraphics[width=88mm]{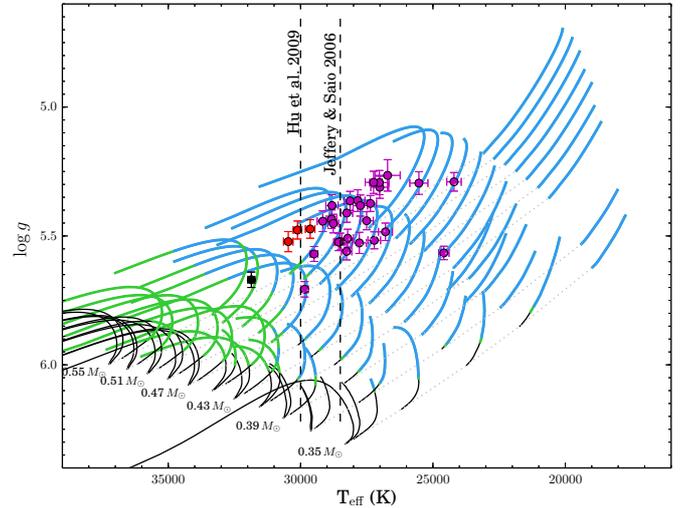}
 \caption{ Same as Fig.\ \ref{sdbgrid_FIG_gmodeInstab}, showing $g$-mode excitation in sdBs but for a grid computed using enhanced envelope mixing. Fewer modes are excited than in the case without extra mixing and it takes longer until enough Fe and Ni is accumulated in the driving region to excite pulsations.}
  \label{sdbgrid_FIG_gmodeInstab_mixing}
\end{figure}

The larger overlap between the $p$- and $g$-mode instability strips compared to previous studies is comforting, since recent work on \emph{Kepler} data has made clear that sdB pulsators that predominantly show $g$-modes often also have one or a few excited $p$-modes \citep[see e.g.][]{BaranKawaler2011} and that $p$-mode pulsator KIC\,10139564 also shows $g$-mode pulsations while it has an effective temperature that is close to 32\,000\,K \citep[][indicated with a black point on Fig.~\ref{sdbgrid_FIG_gmodeInstab}]{BaranReed2012}.  In \citet{HuTout2011} it was shown that a weak stellar wind or a mixing process in the envelope, {effects which are not accounted for in the grid presented in Section~\ref{SEC_instabstrips},} both tend to reduce the build up of Fe and Ni in the envelope. {We have also not considered thermohaline convection \citep[see e.g.][]{Ulrich1972,KippenhahnRuschenplatt1980}, which is another process that could lower the build-up of driving elements by radiative levitation \citep{TheadoVauclair2009}. }

To test the effect of an extra source of turbulent mixing in the envelope, we have recomputed our grid of models using extra turbulent mixing. The resulting instability strip is shown in Fig.~\ref{sdbgrid_FIG_gmodeInstab_mixing}, using the same colours and symbols as in Fig.~\ref{sdbgrid_FIG_gmodeInstab}. Gravity modes are still excited at all temperatures covered by the grid, but fewer modes are excited and the edge of models with {at least 10 excited modes} shifts to $\sim 30\,000\,$K. As a result of the extra mixing it takes longer to build up enough Fe and Ni in the driving region, and models with thin envelopes only start to show excited modes late in their evolution (see black parts of the tracks).

{Our models indicate that when turbulent mixing in the envelope is included, the $g$-mode instability strip still covers all observed $g$-mode pulsators. While this is not a proof that such a mixing process is indeed present, our test shows that we cannot exclude it based on the observed instability strip, which is important because \citet{HuTout2011} showed that surface mixing has the potential to explain the surface He abundances observed in sdBs.}

Stellar evolution calculations with radiative levitation are hampered by
numerical instability because of the short diffusion timescales. With additional turbulent mixing in the outer layer, the numerical stability improves. Therefore, some of the evolution tracks
in Figs.~\ref{sdbgrid_FIG_pmodeInstab} and \ref{sdbgrid_FIG_gmodeInstab} are stopped because of non-convergence, in particular those
of hotter stars that have higher radiative accelerations in the
envelope, while those in Figs.~\ref{sdbgrid_FIG_gmodeInstab_mixing} are more complete.

\section{Conclusions}
We have computed grids of evolutionary subdwarf B star models based on the theoretical work of \citet{HuTout2011} and using their implementation of the atomic diffusion processes, which include radiative levitation, gravitational settling, thermal diffusion and concentration diffusion. The diffusion equations are solved for H, He, C, N, O, Ne, Mg, Fe and Ni. We have measured the build up of Fe and Ni in the pulsation driving region. We find that depending on the total mass of the star and the mass of the hydrogen envelope, Fe gets enhanced by a factor of 10 - 60 and Ni by a factor of 300 - 4000 due to the inclusion of these diffusion processes compared to the initial abundances. {Previous studies proposed that enhancements of Fe and Ni in the driving region around $\log T_{\rm eff} \sim 5.5$ caused pressure and gravity mode pulsations in sdBs. Our results show that these assumptions were valid, and that Fe and Ni are built up to sufficient amounts in the driving region to drive pressure and gravity modes as observed in sdB stars. While previous studies assumed parametrized enhancements of up to a factor of 20, which could explain most of the observed pulsators, we find that the true enhancements factors can be much larger.} We have studied the effect of the Fe and Ni enhancements on the instability strip for $p$- and $g$-mode pulsators by computing the non-adiabatic pulsation properties of the models. While the $g$-mode instability strip in previous studies did not extend to high enough temperatures to include all known pulsators, we obtained a $g$-mode instability strip that predicts that modes can be excited in stars at all effective temperatures covered by the grid, and models have {at least 10 excited modes} at effective temperatures up to $\sim33\,500$\,K, which is \emph{hotter} than what is observed and resolves the problem of the too narrow instability strip. {Further improvements to the work presented here can come from, for example, the inclusion of thermohaline mixing in the models and updates to the opacity tables.} 

\begin{acknowledgements} The research leading to these results has received funding from the
European Research Council under the European Community's Seventh Framework Programme
 FP7-SPACE-2011-1 project n$^\circ$312844 (SPACEINN), as well as from the Research Council of the KU Leuven under grant agreement GOA/2013/012. For the computations we made use of the infrastructure of the VSC -- Flemish Supercomputer Center, funded by the Hercules Foundation and the Flemish Government -- department EWI. During part of this work, SB was supported by the Foundation for Fundamental Research on Matter (FOM),
which is part of the Netherlands Organisation for Scientific Research (NWO). TR is supported by the German Aerospace Center (DLR, grant 05\,OR\,1301).
SB, HH, PD and CA  acknowledge the KITP staff of UCSB for their warm hospitality during the research programme ``Asteroseismology in the Space Age", during which part of this research was performed. This programme was supported by the National Science Foundation, USA, under Grant No. NSF PHY05-51164.
\end{acknowledgements}

\bibliographystyle{aa}
\bibliography{../../../literature/PhD}

\begin{thebibliography}{49}
\expandafter\ifx\csname natexlab\endcsname\relax\def\natexlab#1{#1}\fi

\bibitem[{{Badnell} {et~al.}(2005){Badnell}, {Bautista}, {Butler}, {Delahaye},
  {Mendoza}, {Palmeri}, {Zeippen}, \& {Seaton}}]{BadnellBautista2005}
{Badnell}, N.~R., {Bautista}, M.~A., {Butler}, K., {et~al.} 2005, \mnras, 360,
  458

\bibitem[{{Baran} {et~al.}(2011){Baran}, {Kawaler}, {Reed}, \&
  et~al.}]{BaranKawaler2011}
{Baran}, A.~S., {Kawaler}, S.~D., {Reed}, M.~D., \& et~al. 2011, \mnras, 414,
  2871

\bibitem[{{Baran} {et~al.}(2012){Baran}, {Reed}, {Stello}, \&
  et~al.}]{BaranReed2012}
{Baran}, A.~S., {Reed}, M.~D., {Stello}, D., \& et~al. 2012, \mnras, 424, 2686

\bibitem[{{Beuermann} {et~al.}(2012){Beuermann}, {Dreizler}, {Hessman}, \&
  {Deller}}]{BeuermannDreizler2012}
{Beuermann}, K., {Dreizler}, S., {Hessman}, F.~V., \& {Deller}, J. 2012, \aap,
  543, A138

\bibitem[{{Charpinet} {et~al.}(2009{\natexlab{a}}){Charpinet}, {Brassard},
  {Fontaine}, {Green}, {Van Grootel}, {Randall}, \&
  {Chayer}}]{CharpinetBrassard2009}
{Charpinet}, S., {Brassard}, P., {Fontaine}, G., {et~al.} 2009{\natexlab{a}},
  in American Institute of Physics Conference Series, Vol. 1170, American
  Institute of Physics Conference Series, ed. J.~A. {Guzik} \& P.~A. {Bradley},
  585--596

\bibitem[{{Charpinet} {et~al.}(2009{\natexlab{b}}){Charpinet}, {Fontaine}, \&
  {Brassard}}]{CharpinetFontaine2009}
{Charpinet}, S., {Fontaine}, G., \& {Brassard}, P. 2009{\natexlab{b}}, \aap,
  493, 595

\bibitem[{{Charpinet} {et~al.}(1997){Charpinet}, {Fontaine}, {Brassard},
  {Chayer}, {Rogers}, {Iglesias}, \& {Dorman}}]{CharpinetFontaine1997}
{Charpinet}, S., {Fontaine}, G., {Brassard}, P., {et~al.} 1997, \apjl, 483,
  L123

\bibitem[{{Charpinet} {et~al.}(1996){Charpinet}, {Fontaine}, {Brassard}, \&
  {Dorman}}]{CharpinetFontaine1996}
{Charpinet}, S., {Fontaine}, G., {Brassard}, P., \& {Dorman}, B. 1996, \apjl,
  471, L103

\bibitem[{{Charpinet} {et~al.}(2011){Charpinet}, {Fontaine}, {Brassard},
  {Green}, {Van Grootel}, {Randall}, {Silvotti}, {Baran}, {{\O}stensen},
  {Kawaler}, \& {Telting}}]{CharpinetFontaine2011}
{Charpinet}, S., {Fontaine}, G., {Brassard}, P., {et~al.} 2011, \nat, 480, 496

\bibitem[{{Charpinet} {et~al.}(2013){Charpinet}, {Van Grootel}, {Brassard},
  {Fontaine}, {Green}, \& {Randall}}]{CharpinetVan-Grootel2013}
{Charpinet}, S., {Van Grootel}, V., {Brassard}, P., {et~al.} 2013, in European
  Physical Journal Web of Conferences, Vol.~43, European Physical Journal Web
  of Conferences, 4005

\bibitem[{{Charpinet} {et~al.}(2008){Charpinet}, {Van Grootel}, {Reese},
  {Fontaine}, {Green}, {Brassard}, \& {Chayer}}]{CharpinetVan-Grootel2008}
{Charpinet}, S., {Van Grootel}, V., {Reese}, D., {et~al.} 2008, \aap, 489, 377

\bibitem[{{Dorman} {et~al.}(1993){Dorman}, {Rood}, \&
  {O'Connell}}]{DormanRood1993}
{Dorman}, B., {Rood}, R.~T., \& {O'Connell}, R.~W. 1993, \apj, 419, 596

\bibitem[{{Dupret}(2001)}]{Dupret2001}
{Dupret}, M.~A. 2001, \aap, 366, 166

\bibitem[{{Eggleton}(1971)}]{Eggleton1971}
{Eggleton}, P.~P. 1971, \mnras, 151, 351

\bibitem[{{Fontaine} {et~al.}(2006{\natexlab{a}}){Fontaine}, {Brassard},
  {Charpinet}, \& {Chayer}}]{FontaineBrassard2006a}
{Fontaine}, G., {Brassard}, P., {Charpinet}, S., \& {Chayer}, P.
  2006{\natexlab{a}}, \memsai, 77, 49

\bibitem[{{Fontaine} {et~al.}(2003){Fontaine}, {Brassard}, {Charpinet},
  {Green}, {Chayer}, {Bill{\`e}res}, \& {Randall}}]{FontaineBrassard2003}
{Fontaine}, G., {Brassard}, P., {Charpinet}, S., {et~al.} 2003, \apj, 597, 518

\bibitem[{{Fontaine} {et~al.}(2006{\natexlab{b}}){Fontaine}, {Brassard},
  {Charpinet}, {Green}, {Chayer}, {Randall}, \&
  {Dorman}}]{FontaineBrassard2006}
{Fontaine}, G., {Brassard}, P., {Charpinet}, S., {et~al.} 2006{\natexlab{b}},
  in ESA Special Publication, Vol. 624, Proceedings of SOHO 18/GONG 2006/HELAS
  I, Beyond the spherical Sun

\bibitem[{{Fontaine} {et~al.}(2012){Fontaine}, {Brassard}, {Charpinet},
  {Green}, {Randall}, \& {Van Grootel}}]{FontaineBrassard2012}
{Fontaine}, G., {Brassard}, P., {Charpinet}, S., {et~al.} 2012, \aap, 539, A12

\bibitem[{{Green} {et~al.}(2008){Green}, {Fontaine}, {Hyde}, {For}, \&
  {Chayer}}]{GreenFontaine2008}
{Green}, E.~M., {Fontaine}, G., {Hyde}, E.~A., {For}, B., \& {Chayer}, P. 2008,
  ASPC, 392, 75

\bibitem[{{Green} {et~al.}(2003){Green}, {Fontaine}, {Reed}, {Callerame},
  {Seitenzahl}, {White}, {Hyde}, {{\O}stensen}, {Cordes}, {Brassard}, {Falter},
  {Jeffery}, {Dreizler}, {Schuh}, {Giovanni}, {Edelmann}, {Rigby}, \&
  {Bronowska}}]{GreenFontaine2003}
{Green}, E.~M., {Fontaine}, G., {Reed}, M.~D., {et~al.} 2003, \apjl, 583, L31

\bibitem[{{Grevesse} \& {Noels}(1993)}]{GrevesseNoels1993}
{Grevesse}, N. \& {Noels}, A. 1993, in Origin and Evolution of the Elements,
  ed. N.~{Prantzos}, E.~{Vangioni-Flam}, \& M.~{Casse}, 15--25

\bibitem[{{Han} {et~al.}(2003){Han}, {Podsiadlowski}, {Maxted}, \&
  {Marsh}}]{HanPodsiadlowski2003}
{Han}, Z., {Podsiadlowski}, P., {Maxted}, P.~F.~L., \& {Marsh}, T.~R. 2003,
  \mnras, 341, 669

\bibitem[{{Han} {et~al.}(2002){Han}, {Podsiadlowski}, {Maxted}, {Marsh}, \&
  {Ivanova}}]{HanPodsiadlowski2002}
{Han}, Z., {Podsiadlowski}, P., {Maxted}, P.~F.~L., {Marsh}, T.~R., \&
  {Ivanova}, N. 2002, \mnras, 336, 449

\bibitem[{{Heber}(2009)}]{Heber2009}
{Heber}, U. 2009, \araa, 47, 211

\bibitem[{{Hu} {et~al.}(2008){Hu}, {Dupret}, {Aerts}, {Nelemans}, {Kawaler},
  {Miglio}, {Montalban}, \& {Scuflaire}}]{HuDupret2008}
{Hu}, H., {Dupret}, M., {Aerts}, C., {et~al.} 2008, \aap, 490, 243

\bibitem[{{Hu} {et~al.}(2010){Hu}, {Glebbeek}, {Thoul}, {Dupret}, {Stancliffe},
  {Nelemans}, \& {Aerts}}]{HuGlebbeek2010}
{Hu}, H., {Glebbeek}, E., {Thoul}, A.~A., {et~al.} 2010, \aap, 511, A87+

\bibitem[{{Hu} {et~al.}(2009){Hu}, {Nelemans}, {Aerts}, \&
  {Dupret}}]{HuNelemans2009}
{Hu}, H., {Nelemans}, G., {Aerts}, C., \& {Dupret}, M. 2009, \aap, 508, 869

\bibitem[{{Hu} {et~al.}(2011){Hu}, {Tout}, {Glebbeek}, \&
  {Dupret}}]{HuTout2011}
{Hu}, H., {Tout}, C.~A., {Glebbeek}, E., \& {Dupret}, M.-A. 2011, \mnras, 418,
  195

\bibitem[{{Iglesias} \& {Rogers}(1996)}]{IglesiasRogers1996}
{Iglesias}, C.~A. \& {Rogers}, F.~J. 1996, \apj, 464, 943

\bibitem[{{Jeffery} \& {Saio}(2006)}]{JefferySaio2006}
{Jeffery}, C.~S. \& {Saio}, H. 2006, \mnras, 372, L48

\bibitem[{{Kawaler}(2010)}]{Kawaler2010}
{Kawaler}, S.~D. 2010, Astronomische Nachrichten, 331, 1020

\bibitem[{{Kawaler} \& {Hostler}(2005)}]{KawalerHostler2005}
{Kawaler}, S.~D. \& {Hostler}, S.~R. 2005, \apj, 621, 432

\bibitem[{{Kilkenny} {et~al.}(1997){Kilkenny}, {Koen}, {O'Donoghue}, \&
  {Stobie}}]{KilkennyKoen1997}
{Kilkenny}, D., {Koen}, C., {O'Donoghue}, D., \& {Stobie}, R.~S. 1997, \mnras,
  285, 640

\bibitem[{{Kippenhahn} {et~al.}(1980){Kippenhahn}, {Ruschenplatt}, \&
  {Thomas}}]{KippenhahnRuschenplatt1980}
{Kippenhahn}, R., {Ruschenplatt}, G., \& {Thomas}, H.-C. 1980, \aap, 91, 175

\bibitem[{{Maxted} {et~al.}(2001){Maxted}, {Heber}, {Marsh}, \&
  {North}}]{MaxtedHeber2001}
{Maxted}, P.~f.~L., {Heber}, U., {Marsh}, T.~R., \& {North}, R.~C. 2001,
  \mnras, 326, 1391

\bibitem[{{Michaud} {et~al.}(2011){Michaud}, {Richer}, \&
  {Richard}}]{MichaudRicher2011}
{Michaud}, G., {Richer}, J., \& {Richard}, O. 2011, \aap, 529, A60

\bibitem[{{{\O}stensen}(2009)}]{Ostensen2009}
{{\O}stensen}, R.~H. 2009, Communications in Asteroseismology, 159, 75

\bibitem[{{{\O}stensen}(2010)}]{Ostensen2010}
{{\O}stensen}, R.~H. 2010, Astronomische Nachrichten, 331, 1026

\bibitem[{{Pablo} {et~al.}(2011){Pablo}, {Kawaler}, \&
  {Green}}]{PabloKawaler2011}
{Pablo}, H., {Kawaler}, S.~D., \& {Green}, E.~M. 2011, \apjl, 740, L47

\bibitem[{{Pablo} {et~al.}(2012){Pablo}, {Kawaler}, {Reed}, \&
  et~al.}]{PabloKawaler2012}
{Pablo}, H., {Kawaler}, S.~D., {Reed}, M.~D., \& et~al. 2012, \mnras, 422, 1343

\bibitem[{{Podsiadlowski} {et~al.}(2008){Podsiadlowski}, {Han}, {Lynas-Gray},
  \& {Brown}}]{PodsiadlowskiHan2008}
{Podsiadlowski}, P., {Han}, Z., {Lynas-Gray}, A.~E., \& {Brown}, D. 2008, in
  Astronomical Society of the Pacific Conference Series, Vol. 392, Hot Subdwarf
  Stars and Related Objects, ed. {U.~Heber, C.~S.~Jeffery, \& R.~Napiwotzki},
  15

\bibitem[{{Silvotti} {et~al.}(2007){Silvotti}, {Schuh}, {Janulis}, \&
  et~al.}]{SilvottiSchuh2007}
{Silvotti}, R., {Schuh}, S., {Janulis}, R., \& et~al. 2007, \nat, 449, 189

\bibitem[{{Soker} \& {Harpaz}(2000)}]{SokerHarpaz2000}
{Soker}, N. \& {Harpaz}, A. 2000, \mnras, 317, 861

\bibitem[{{Th{\'e}ado} {et~al.}(2009){Th{\'e}ado}, {Vauclair}, {Alecian}, \&
  {Le Blanc}}]{TheadoVauclair2009}
{Th{\'e}ado}, S., {Vauclair}, S., {Alecian}, G., \& {Le Blanc}, F. 2009, \apj,
  704, 1262

\bibitem[{{Ulrich}(1972)}]{Ulrich1972}
{Ulrich}, R.~K. 1972, \apj, 172, 165

\bibitem[{{Unno} {et~al.}(1989){Unno}, {Osaki}, {Ando}, {Saio}, \&
  {Shibahashi}}]{UnnoOsaki1989}
{Unno}, W., {Osaki}, Y., {Ando}, H., {Saio}, H., \& {Shibahashi}, H. 1989,
  {Nonradial oscillations of stars}

\bibitem[{{Van Grootel} {et~al.}(2013){Van Grootel}, {Charpinet}, {Brassard},
  {Fontaine}, \& {Green}}]{Van-GrootelCharpinet2013}
{Van Grootel}, V., {Charpinet}, S., {Brassard}, P., {Fontaine}, G., \& {Green},
  E.~M. 2013, \aap, 553, A97

\bibitem[{{Van Grootel} {et~al.}(2008){Van Grootel}, {Charpinet}, {Fontaine},
  \& {Brassard}}]{Van-GrootelCharpinet2008}
{Van Grootel}, V., {Charpinet}, S., {Fontaine}, G., \& {Brassard}, P. 2008,
  \aap, 483, 875

\bibitem[{{Van Grootel} {et~al.}(2010){Van Grootel}, {Charpinet}, {Fontaine},
  {Brassard}, {Green}, {Randall}, {Silvotti}, {{\O}stensen}, {Kjeldsen},
  {Christensen-Dalsgaard}, {Borucki}, \& {Koch}}]{Van-GrootelCharpinet2010}
{Van Grootel}, V., {Charpinet}, S., {Fontaine}, G., {et~al.} 2010, \apjl, 718,
  L97

\end{thebibliography}

\begin{appendix}
\section{Excited modes for different spherical degrees}\label{ch_appendix_047}
{
In this Appendix, we provide detailed information on the model with a total mass of $0.47$M$_\odot$ and an envelope mass of $9.0\ 10^{-5}$M$_\odot$, for which the pulsational characteristics were shown in Fig.\ \ref{fig_0.47pgmodeVSl}. In Tables \ref{tab_0.47pmode} and \ref{tab_0.47gmode} the period range and the number of excited modes is given for $p$- and $g$-modes, respectively.  The tables also list the effective temperature (T$_{\rm eff}$), surface gravity ($\log g$) and Fe and Ni abundance enhancements (X$_{\rm Fe}$/X$_{\rm Fe,init}$ and X$_{\rm Ni}$/X$_{\rm Ni,init}$) in the driving region as a function of time.}

\begin{table*}
\caption{{Number (no.), period range ($p$), and order range ($n$) of excited $p$-modes for different spherical degrees for the evolutionary track with a total mass of $0.47$M$_\odot$ and an envelope mass of $9.0\ 10^{-5}$M$_\odot$.}}              
\label{tab_0.47pmode}      
\centering                                      

\tabcolsep=2.7pt

\begin{tabular}{c c c c c | c c c | c c c | c c c | c c c | c c c}          
\hline\hline                        

T$_{\rm eff}$  & $\log g$ & Age  & X$_{\rm Fe}$ & X$_{\rm Ni}$ & \multicolumn{3}{l}{$l=0$} & \multicolumn{3}{l}{$l=1$} & \multicolumn{3}{l}{$l=2$} & \multicolumn{3}{l}{$l=3$} & \multicolumn{3}{l}{$l=4$} \\ 
 (K) & & (Myr) & /X$_{\rm Fe,init}$ & /X$_{\rm Ni,init}$ & no. & p$_{\rm range}$ (s) & n$_{\rm range}$ & no. & p$_{\rm range}$ (s) & n$_{\rm range}$ & no. & p$_{\rm range}$ (s) & n$_{\rm range}$ & no. & p$_{\rm range}$ (s) & n$_{\rm range}$ & no. & p$_{\rm range}$ (s) & n$_{\rm range}$ \\
\hline
34142  & 6.07  &   0.0 &  1.0 &  1.0 & 0 & ---  & --- & 0 & ---  & --- & 0 & ---  & --- & 0 & ---  & --- & 0 & ---  & --- \\
32769  & 5.99  &   5.1 & 33.8 & 1602.6 &   6 & 68--149  &  1-- 6 &   6 & 65--132  &  1-- 6 &   6 & 62--120  &  1-- 6 &   5 & 67--112  &  1-- 5 &   5 & 65--107  &  1-- 5 \\
32544  & 5.97  &  10.4 & 34.0 & 1658.2 &   6 & 72--151  &  1-- 6 &   6 & 68--134  &  1-- 6 &   5 & 72--123  &  1-- 5 &   5 & 70--114  &  1-- 5 &   5 & 67--109  &  1-- 5 \\
32413  & 5.96  &  14.5 & 33.1 & 1636.9 &   6 & 73--153  &  1-- 6 &   6 & 69--136  &  1-- 6 &   5 & 74--124  &  1-- 5 &   5 & 71--116  &  1-- 5 &   5 & 69--111  &  1-- 5 \\
32311  & 5.95  &  19.7 & 33.8 & 1678.9 &   6 & 75--155  &  1-- 6 &   6 & 70--138  &  1-- 6 &   5 & 76--126  &  1-- 5 &   5 & 72--118  &  1-- 5 &   4 & 70--114  &  1-- 5 \\
32240  & 5.94  &  25.1 & 31.4 & 1625.7 &   6 & 76--157  &  1-- 6 &   6 & 71--140  &  1-- 6 &   5 & 77--128  &  1-- 5 &   5 & 74--120  &  1-- 5 &   5 & 71--116  &  1-- 5 \\
32185  & 5.93  &  30.5 & 32.1 & 1670.8 &   6 & 77--158  &  1-- 6 &   6 & 72--142  &  1-- 6 &   5 & 78--130  &  1-- 5 &   5 & 75--122  &  1-- 5 &   5 & 72--118  &  1-- 5 \\
32159  & 5.92  &  34.8 & 34.3 & 1776.4 &   6 & 78--160  &  1-- 6 &   5 & 84--144  &  1-- 5 &   5 & 79--131  &  1-- 5 &   5 & 76--123  &  1-- 5 &   4 & 84--120  &  1-- 4 \\
32134  & 5.91  &  40.1 & 34.7 & 1821.1 &   6 & 80--162  &  1-- 6 &   5 & 85--146  &  1-- 5 &   5 & 81--133  &  1-- 5 &   5 & 77--125  &  1-- 5 &   4 & 85--121  &  1-- 4 \\
32102  & 5.90  &  45.1 & 33.6 & 1793.2 &   6 & 81--163  &  1-- 6 &   6 & 76--148  &  1-- 6 &   5 & 82--135  &  1-- 5 &   5 & 78--127  &  1-- 5 &   4 & 87--123  &  1-- 4 \\
32092  & 5.89  &  50.4 & 35.2 & 1895.1 &   6 & 82--165  &  1-- 6 &   5 & 88--150  &  1-- 5 &   5 & 83--137  &  1-- 5 &   5 & 79--128  &  1-- 5 &   4 & 88--125  &  1-- 4 \\
32072  & 5.88  &  54.7 & 34.1 & 1863.1 &   6 & 83--166  &  1-- 6 &   6 & 78--151  &  1-- 6 &   5 & 84--138  &  1-- 5 &   5 & 80--130  &  1-- 5 &   4 & 89--127  &  1-- 4 \\
32070  & 5.87  &  60.1 & 36.3 & 1988.1 &   6 & 84--168  &  1-- 6 &   5 & 91--153  &  1-- 5 &   5 & 85--140  &  1-- 5 &   5 & 81--132  &  1-- 5 &   4 & 91--129  &  1-- 4 \\
32058  & 5.87  &  65.4 & 32.8 & 1874.3 &   6 & 85--170  &  1-- 6 &   5 & 92--156  &  1-- 5 &   5 & 87--142  &  1-- 5 &   5 & 83--134  &  1-- 5 &   4 & 92--131  &  1-- 4 \\
32052  & 5.86  &  69.6 & 35.1 & 1988.5 &   6 & 87--171  &  1-- 6 &   5 & 93--157  &  1-- 5 &   5 & 88--144  &  1-- 5 &   5 & 84--135  &  1-- 5 &   4 & 93--132  &  1-- 4 \\
32059  & 5.85  &  74.9 & 37.0 & 2101.0 &   6 & 88--173  &  1-- 6 &   5 & 95--159  &  1-- 5 &   5 & 89--145  &  1-- 5 &   5 & 85--137  &  1-- 5 &   4 & 95--135  &  1-- 4 \\
32071  & 5.84  &  80.1 & 35.1 & 2062.7 &   6 & 89--175  &  1-- 6 &   5 & 96--162  &  1-- 5 &   6 & 36--147  &  1--17 &   5 & 86--139  &  1-- 5 &   4 & 96--137  &  1-- 4 \\
32072  & 5.83  &  85.3 & 36.3 & 2135.2 &   6 & 90--177  &  1-- 6 &   5 & 97--164  &  1-- 5 &   6 & 37--149  &  1--17 &   5 & 88--141  &  1-- 5 &   4 & 98--139  &  1-- 4 \\
32089  & 5.82  &  89.5 & 37.7 & 2229.6 &   6 & 92--179  &  1-- 6 &   5 & 98--165  &  1-- 5 &   6 & 37--151  &  1--17 &   5 & 89--143  &  1-- 5 &   4 & 99--140  &  1-- 4 \\
32116  & 5.82  &  94.9 & 37.6 & 2259.1 &   6 & 93--180  &  1-- 6 &   5 & 100--168  &  1-- 5 &   6 & 38--153  &  1--17 &   5 & 90--145  &  1-- 5 &   4 & 100--143  &  1-- 4 \\
32145  & 5.81  & 100.2 & 37.7 & 2312.2 &   6 & 94--182  &  1-- 6 &   5 & 101--170  &  1-- 5 &   6 & 39--155  &  1--17 &   5 & 91--147  &  1-- 5 &   5 & 39--145  &  1--16 \\
32182  & 5.80  & 105.4 & 37.6 & 2353.1 &   6 & 96--184  &  1-- 6 &   6 & 40--172  &  1--17 &   6 & 39--156  &  1--17 &   5 & 93--149  &  1-- 5 &   5 & 40--147  &  1--16 \\
32215  & 5.79  & 109.5 & 37.8 & 2391.8 &   6 & 97--186  &  1-- 6 &   6 & 40--174  &  1--17 &   6 & 40--158  &  1--17 &   5 & 94--151  &  1-- 5 &   5 & 40--149  &  1--16 \\
32258  & 5.78  & 114.6 & 40.2 & 2540.6 &   6 & 98--187  &  1-- 6 &   5 & 105--176  &  1-- 5 &   5 & 99--160  &  1-- 5 &   5 & 95--153  &  1-- 5 &   5 & 41--151  &  1--16 \\
32327  & 5.78  & 119.7 & 42.7 & 2744.9 &   6 & 99--189  &  1-- 6 &   6 & 42--178  &  1--17 &   5 & 100--161  &  1-- 5 &   5 & 96--155  &  1-- 5 &   5 & 41--153  &  1--16 \\\hline                                             
\end{tabular}
\end{table*}

\begin{table*}
\caption{{Same as Table \ref{tab_0.47pmode} but for excited $g$-modes.}}              
\label{tab_0.47gmode}      
\centering                                      

\tabcolsep=4.35pt

\begin{tabular}{c c c c c |  c c c | c c c | c c c | c c c}          
\hline\hline                        

T$_{\rm eff}$  & $\log g$ & Age  & X$_{\rm Fe}$ & X$_{\rm Ni}$ & \multicolumn{3}{l}{$l=1$} & \multicolumn{3}{l}{$l=2$} & \multicolumn{3}{l}{$l=3$} & \multicolumn{3}{l}{$l=4$} \\ 
 (K) & & (Myr) & /X$_{\rm Fe,init}$ & /X$_{\rm Ni,init}$ & no. & p$_{\rm range}$ (s) & n$_{\rm range}$ & no. & p$_{\rm range}$ (s) & n$_{\rm range}$ & no. & p$_{\rm range}$ (s) & n$_{\rm range}$ & no. & p$_{\rm range}$ (s) & n$_{\rm range}$ \\
\hline
34142  & 6.07  &   0.0 &  1.0 &  1.0 & 0 & ---  & --- & 0 & ---  & --- & 0 & ---  & --- & 0 & ---  & --- \\
32769  & 5.99  &   5.1 & 33.8 & 1602.6 & 0 & ---  & --- &   7 & 298--1433  &  1--10 &  11 & 237--1324  &  1--14 &  14 & 207--1241  &  1--17 \\
32544  & 5.97  &  10.4 & 34.0 & 1658.2 & 0 & ---  & --- &   9 & 298--1640  &  1--12 &  14 & 238--1541  &  1--17 &  18 & 208--1359  &  1--20 \\
32413  & 5.96  &  14.5 & 33.1 & 1636.9 & 0 & ---  & --- &   8 & 297--1672  &  1--12 &  15 & 238--1578  &  1--18 &  17 & 209--1335  &  1--20 \\
32311  & 5.95  &  19.7 & 33.8 & 1678.9 & 0 & ---  & --- &   9 & 297--1775  &  1--13 &  17 & 238--4877  &  1--55 &  18 & 209--1443  &  1--21 \\
32240  & 5.94  &  25.1 & 31.4 & 1625.7 & 0 & ---  & --- &   9 & 296--1773  &  1--13 &  16 & 237--5048  &  1--56 &  17 & 209--1478  &  1--21 \\
32185  & 5.93  &  30.5 & 32.1 & 1670.8 & 0 & ---  & --- &  10 & 295--1959  &  1--14 &  15 & 236--5066  &  1--56 &  18 & 208--1508  &  1--21 \\
32159  & 5.92  &  34.8 & 34.3 & 1776.4 & 0 & ---  & --- &  10 & 295--1988  &  1--14 &  16 & 236--5251  &  1--58 &  18 & 208--4057  &  1--58 \\
32134  & 5.91  &  40.1 & 34.7 & 1821.1 & 0 & ---  & --- &  10 & 294--2017  &  1--14 &  17 & 235--5230  &  1--57 &  20 & 207--1628  &  1--23 \\
32102  & 5.90  &  45.1 & 33.6 & 1793.2 & 0 & ---  & --- &  10 & 293--2041  &  1--14 &  17 & 234--5057  &  1--55 &  20 & 206--1646  &  1--23 \\
32092  & 5.89  &  50.4 & 35.2 & 1895.1 & 0 & ---  & --- &  11 & 293--2099  &  1--15 &  17 & 234--2279  &  1--23 &  21 & 206--3988  &  1--56 \\
32072  & 5.88  &  54.7 & 34.1 & 1863.1 & 0 & ---  & --- &  11 & 292--2116  &  1--15 &  17 & 233--5012  &  1--54 &  21 & 205--1689  &  1--24 \\
32070  & 5.87  &  60.1 & 36.3 & 1988.1 &   1 & 455  &  1 &  11 & 292--2143  &  1--15 &  17 & 233--1977  &  1--20 &  22 & 204--1774  &  1--25 \\
32058  & 5.87  &  65.4 & 32.8 & 1874.3 &   1 & 454  &  1 &  11 & 291--2171  &  1--15 &  17 & 232--1931  &  1--20 &  22 & 204--1795  &  1--25 \\
32052  & 5.86  &  69.6 & 35.1 & 1988.5 &   1 & 454  &  1 &  12 & 291--2192  &  1--15 &  19 & 231--4955  &  1--53 &  21 & 203--1811  &  1--25 \\
32059  & 5.85  &  74.9 & 37.0 & 2101.0 &   1 & 454  &  1 &  13 & 290--2350  &  1--16 &  18 & 231--1965  &  1--21 &  22 & 202--1850  &  1--26 \\
32071  & 5.84  &  80.1 & 35.1 & 2062.7 &   1 & 453  &  1 &  15 & 290--7120  &  1--53 &  17 & 230--1988  &  1--21 &  22 & 202--1861  &  1--26 \\
32072  & 5.83  &  85.3 & 36.3 & 2135.2 &   1 & 452  &  1 &  12 & 289--2287  &  1--16 &  18 & 230--2123  &  1--22 &  22 & 201--1881  &  1--26 \\
32089  & 5.82  &  89.5 & 37.7 & 2229.6 &   1 & 452  &  1 &  13 & 289--2424  &  1--17 &  19 & 229--2163  &  1--23 &  22 & 201--1896  &  1--26 \\
32116  & 5.82  &  94.9 & 37.6 & 2259.1 &   1 & 451  &  1 &  14 & 288--2534  &  1--18 &  22 & 228--4896  &  1--52 &  24 & 200--2024  &  1--28 \\
32145  & 5.81  & 100.2 & 37.7 & 2312.2 &   1 & 450  &  1 &  14 & 287--2557  &  1--18 &  20 & 228--2321  &  1--24 &  24 & 199--2043  &  1--28 \\
32182  & 5.80  & 105.4 & 37.6 & 2353.1 &   1 & 449  &  1 &  16 & 287--2635  &  1--20 &  19 & 227--2358  &  1--24 &  25 & 199--3637  &  1--48 \\
32215  & 5.79  & 109.5 & 37.8 & 2391.8 &   1 & 448  &  1 &  17 & 286--6561  &  1--49 &  21 & 226--4716  &  1--50 &  25 & 198--2154  &  1--29 \\
32258  & 5.78  & 114.6 & 40.2 & 2540.6 &   2 & 446--2459  &  1-- 9 &  15 & 285--2625  &  1--19 &  23 & 226--4608  &  1--48 &  27 & 198--2197  &  1--30 \\
32327  & 5.78  & 119.7 & 42.7 & 2744.9 &   6 & 444--3058  &  1--12 &  19 & 284--6805  &  1--51 &  21 & 225--2444  &  1--25 &  28 & 198--2338  &  1--31 \\
\hline                                             
\end{tabular}
\end{table*}

\end{appendix}
\end{document}